\documentclass{IEEEtran}
\ifCLASSINFOpdf
\else
   \usepackage[dvips]{graphicx}
\fi
\usepackage{url}

\usepackage{graphicx}
\usepackage{amsmath,hyperref,amssymb,amsfonts}
\usepackage{textcomp}
\usepackage{xcolor}
\usepackage{tikz}
\usepackage{array}
\usepackage{multirow}
\usepackage{booktabs}
\usepackage{makecell}
\usepackage{booktabs}
\usepackage{algorithm,algorithmic}
\usepackage{verbatim}

\begin{document}

\title{Noise-Conditioned Mixture-of-Experts Framework for Robust Speaker Verification}

\author{Bin Gu, \IEEEmembership{Member, IEEE}, Haitao Zhao, \IEEEmembership{Senior Member, IEEE}, Jibo Wei, \IEEEmembership{Member, IEEE}
\thanks{This paragraph of the first footnote will contain the date on which you submitted your paper for review. This work was partially funded by the Provincial Department of Education Program (Grant No. 25B0597). }
\thanks{Bin Gu, Haitao Zhao and Jibo Wei are with the College of Electronic Science and Technology, National University of Defense Technology, Changsha, China (e-mail: gb24@nudt.edu.cn; haitaozhao@nudt.edu.cn;wjbhw@nudt.edu.cn).}}

\markboth{Journal of \LaTeX\ Class Files, Vol. 14, No. 8, August 2015}
{Shell \MakeLowercase{\textit{et al.}}: Bare Demo of IEEEtran.cls for IEEE Journals}
\maketitle

\begin{abstract}
Robust speaker verification under noisy conditions remains an open challenge. Conventional deep learning methods learn a robust unified speaker representation space against diverse background noise and achieve significant improvement. In contrast, this paper presents a noise-conditioned mixture-of-experts framework that decomposes the feature space into specialized noise-aware subspaces for speaker verification. Specifically, we propose a noise-conditioned expert routing mechanism, a universal model based expert specialization strategy, and an SNR-decaying curriculum learning  protocol, collectively improving model robustness and generalization under diverse noise conditions. The proposed method can automatically route inputs to expert networks based on noise information derived from the inputs, where each expert targets distinct noise characteristics while preserving speaker identity information. Comprehensive experiments demonstrate consistent superiority over baselines.
\end{abstract}


\IEEEpeerreviewmaketitle

\vspace{-0.5cm}
\section{Introduction}

\IEEEPARstart{S}{peaker} verification (SV), which aims to verify the identity of a given utterance\cite{b1}, has been widely adopted in smart devices and other security-critical applications. While deep learning has substantially advanced SV systems \cite{b2,b3,b5,b27,b6}, their real-world deployment still faces significant challenges, as the shift to unconstrained environments brings challenging acoustic interference \cite{b7}. Common noise sources, including ambient music, non-stationary noise, and crowd babble, create diverse spectral distortions that substantially degrade verification performance. This robustness challenge continues to hinder reliable real-world implementation. 

To alleviate the above-mentioned problem, a dominant and effective approach is to employ speech enhancement (SE) networks for noise suppression, with the enhanced feature then utilized for SV \cite{b13,b14}. Compared to cascading pre-trained SE and SV networks, \cite{b8} showed that constructing an SE model specialized for SV tasks yields better performance. Thereafter, some studies have attempted to boost system robustness through targeted improvements to SE modules. In \cite{b9}, both masking- and mapping-based SE networks are integrated into the SV system to remove noise from different aspects. In \cite{b10}, a novel extened U-Net is adopted as the SE backbone and showed the superiority of the jointly optimized cascade system through end-to-end learning. To mitigate error accumulation, \cite{b33} aggregates lost information for speaker restoration, while \cite{b34} preserves speaker information via explicit noise modeling with dual U-Nets. With the advent of diffusion models, recent works \cite{b11}\cite{b12} have incorporated them into the SE front-end, achieving notable performance gains. 

An alternative technical paradigm employs advanced learning strategies to derive noise-invariant speaker representations. As demonstrated in \cite{b15}, feeding clean and noisy speech into the network while minimizing the distances at the embedding level can improve robustness. Further improvements are achieved by \cite{b12} through supervised contrastive learning to narrow distribution gaps between clean and noisy samples, and by \cite{b16} via disentanglement methods to isolate noise components from speech representations. \cite{b17} extends this framework by incorporating adversarial training to treat different noise types as distinct domains, making their representations domain indistinguishable. Additionally, \cite{b18} introduces stable learning to eliminate spurious correlations in training data, thereby improving noise generalization. 

\begin{figure*}[t]
\centerline{\includegraphics[width=0.9\linewidth]{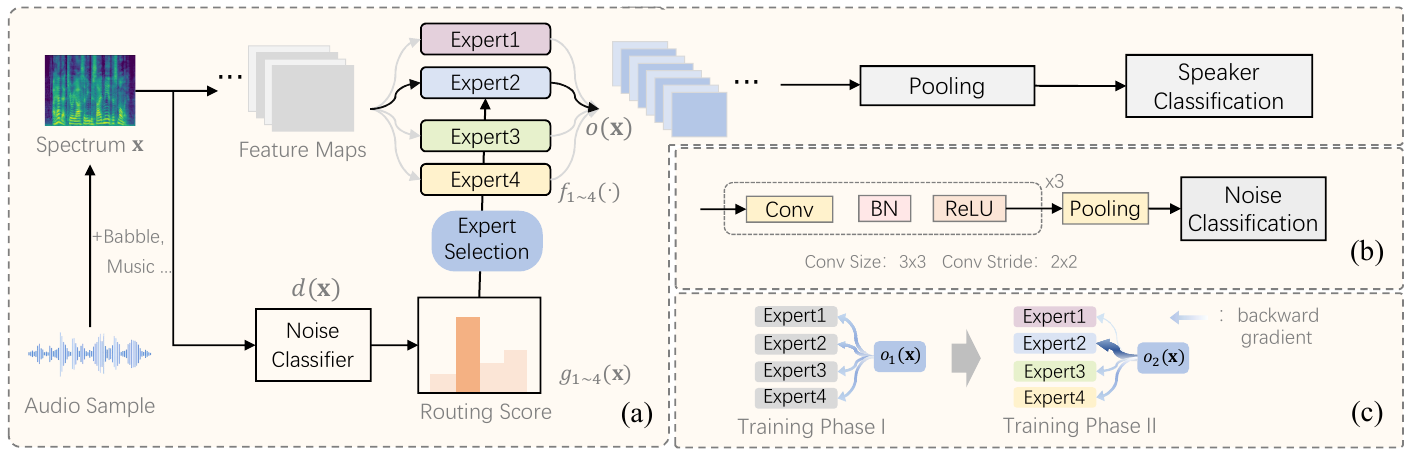}}
\vspace{-0.2cm}
\caption{(a) The noise-conditioned mixture-of-expert framework, (b) the noise classifier architecture, and (c) the gradient flow in UMES.}
\vspace{-0.5cm}
\label{fig1}
\end{figure*}

While both kinds of approaches have shown promising results, their dependence on unified feature-space modeling may present certain constraints. In cases where input distributions exhibit notable variations, maintaining effective discrimination within a single feature space could prove challenging. Mixture-of-experts (MoE) framework \cite{b31,b32}, which leverage multiple specialized sub-networks, have demonstrated strong performance across various tasks by enabling more flexible feature representations \cite{b25}. However, their potential for improving robustness in speaker verification remains underexplored. Building upon this, we explore a noise-conditioned mixture-of-expert (NCMoE) framework and investigate the feasibility of decomposing the feature space into noise-specific subspaces. This exploration into condition-specific subspace modeling is intended to offer a potential alternative for speaker feature extraction in challenging environments, and we hope it may provide useful insights for robust speaker representation learning. Specifically, a lightweight convolutional network first estimates the noise characteristics of spectral features to guide the selection of specialized expert branches. Subsequently, to facilitate expert training, we introduce a universal model-based strategy. This strategy begins with a generalist model and then progressively specializes it into noise-adapted subspaces, a process further stabilized by an SNR-progressive curriculum. Experiments on VoxCeleb1 with simulated noise conditions suggest this approach outperforms conventional baselines. 

\section{Noise-Conditioned Mixture-of-Expert Framework}

\subsection{Overview of the Framework}

As illustrated in Fig.\ref{fig1} (a), our framework preserves the backbone network's original architecture while augmenting a selected intermediate layer with parallel expert branches to balance model capacity and efficiency. Each expert replicates the complete structure of the original layer (i.e., maintaining identical input/output dimensions and internal connections), ensuring architectural consistency. In addition, a compact noise classification network (Fig.\ref{fig1} (b)) dynamically selects a single expert branch per sample during forward propagation, keeping others inactive. This design ensures computational efficiency through sparse expert activation while maintaining manageable resource requirements, allowing specialized processing tailored to distinct noise conditions. 
\vspace{-0.3cm}
\subsection{Noise-Conditioned Expert Routing}
The Noise-Conditioned Expert Routing (NCER) method dynamically selects processing paths based on input features. Given an input feature $\mathbf{x} \in \mathbb{R}^{F \times T}$, the noise classifier $d(\mathbf{x})$ predicts the noise category distribution $\hat{\mathbf{y}} = [g_1(\mathbf{x}), \dots, g_n(\mathbf{x})]$, where each routing value $g_i(\mathbf{x})$ for routing expert is computed via temperature-scaled softmax:

\begin{equation}
g_i(\mathbf{x}) = \frac{\exp(z_i/\gamma)}{\sum_{j=1}^n \exp(z_j/\gamma)}.
\end{equation}
The classifier $d(\cdot)$ consists of sequential strided convolutions followed by temporal pooling and softmax, with $z_i$ denoting the logit for noise class $i$ and temperature factor $\gamma$ controlling output sharpness. As shown in Fig.\ref{fig1} (a), the MoE's output $o(\mathbf{x})$ is determined by a gated selection of experts, which operates differently during training and testing:
\vspace{-0.1cm}
\begin{equation}
o(\mathbf{x}) = \begin{cases}
\sum_{i=1}^n g_i(\mathbf{x}) f_i(\mathbf{x}) & \text{if training} \\
f_{i}(\mathbf{x}), \quad i = \mathop{\mathrm{arg\,max}}\limits_{k} g_k(\mathbf{x}) & \text{if testing}
\end{cases}
\label{eq2}
\end{equation}
where ${\{f_i(\cdot)\}}_{i=1}^n$ are expert networks. The conditional execution strategy in Eq.\ref{eq2} ensures proper gradient flow through all experts during backpropagation (training phase), while eliminating unnecessary computational overhead from inactive branches during inference (testing phase). Note that all experts share identical architectures, and the noise classifier uses only three convolutional layers. This simplicity ensures performance gains stem solely from the routing strategy, not network architecture design.

\begin{table*}[t!]
\footnotesize
\centering
\caption{RESULTS (EER\%) ON VOXCELEB1 TEST SET WITH MUSAN DATA AT VARIOUS SNRS}
\vspace{-0.3cm}
\label{tab:results}
\resizebox{\textwidth}{!}{
\begin{tabular}{c|c|ccccccccccc}
\cline{1-13}
\toprule
\multicolumn{2}{c|}{Training Set}&\multicolumn{11}{c}{VoxCeleb1}\\

\cline{1-13}
Noise Type& SNR &Baseline& VoiceID\textsuperscript{\cite{b8}}  & FSEF\textsuperscript{\cite{b9}}  & NDML\textsuperscript{\cite{b16}}  & WSVIL\textsuperscript{\cite{b15}}  & ExU-Net\textsuperscript{\cite{b10}}  & SEU-Net\textsuperscript{\cite{b18}}& Diff-SV\textsuperscript{\cite{b11}}  & NDAL\textsuperscript{\cite{b17}}  &NISRL\textsuperscript{\cite{b12}}  & NCMoE \\
\cline{1-13}

\multicolumn{2}{c|}{Original Set} & 1.98 & 6.79 & 4.26 & 2.90 & 3.12 & 2.76& 2.52 & 2.35 & 2.63 & 2.40 & \textbf{1.91} \\
\cline{1-13}
\multirow{5}{*}{Babble} 
& 0 & 9.30 & 38.0 & 27.6 & 11.0 & 11.8 & 9.57 &8.54 & 8.74 & \textbf{6.43} & 7.81 &8.10\\
& 5 & 4.56 & 27.1 & 15.3 & 6.13 & 5.97 & 5.52 &5.16 & 4.51 & 4.44 & 4.25 &\textbf{4.24}\\
& 10 & 2.99 & 16.7 & 9.04 & 4.28 & 4.44 & 4.06 &3.67 & 3.33 & 3.59 & 3.28 &\textbf{2.88}\\
& 15 & 2.45 & 11.3 & 6.47 & 3.52 & 3.73 & 3.28 &3.10 & 2.82 & 3.08 & 2.78 &\textbf{2.51}\\
& 20 & 2.18 & 8.99 & 5.41 & 3.21 & 3.36 & 2.99 &2.79 & 2.61 & 2.87 & 2.60 &\textbf{2.06}\\
\cline{1-13}

\multirow{5}{*}{Music}
& 0 & 5.82 & 16.2 & 8.47 & 10.8 & 7.79 & 7.35 &6.25 & 6.04 & 5.87 & 5.19 &\textbf{4.62}\\
& 5 & 3.57 & 11.4 & 6.31 & 6.52 & 5.23 & 4.90 &4.36 & 3.96 & 4.19 & 3.58 &\textbf{3.04}\\
& 10 & 2.73 & 9.13 & 5.14 & 4.66 & 4.11 & 3.69 &3.55 & 3.10 & 3.53 & 3.11 &\textbf{2.50}\\
& 15 & 2.28 & 8.10 & 4.71 & 3.67 & 3.63 & 3.14  &3.10 & 2.75 & 3.23& 2.75 &\textbf{2.11}\\
& 20 & 2.13 & 7.48 & 4.56 & 3.21 & 3.30 & 2.93 &2.79 & 2.60 & 3.09 & 2.57 &\textbf{2.07}\\
\cline{1-13}

\multirow{5}{*}{Noise}
& 0 & 7.3 & 16.6 & 7.88 & 10.2 & 7.34 & 6.80 &6.41 & 6.01 & 6.14 & \textbf{4.94} &5.20\\
& 5 & 4.45 & 12.3 & 6.42 & 6.96 & 5.65 & 5.23 &4.42 & 4.52 & 4.00 & 3.69 &\textbf{3.65}\\
& 10 & 3.14 & 9.86 & 5.50 & 5.02 & 4.35 & 4.07 &3.74 & 3.49 & 3.23 & 3.43 &\textbf{2.72}\\
& 15 & 2.57 & 8.69 & 4.87 & 3.91 & 3.85 & 3.39 &3.20 & 2.93 & 2.97 & 2.94 &\textbf{2.39}\\
& 20 & 2.25 & 7.83 & 4.66 & 3.40 & 3.44 & 3.10 &2.92 & 2.64 & 2.80 & 2.68 &\textbf{2.16}\\
\cline{1-13}

\multicolumn{2}{c|}{Average} & 3.73& 13.5 & 7.91 & 5.59 & 5.07 & 4.55 &4.16  & 3.90 & 3.88 & 3.62 & \textbf{3.26}\\
\cline{1-13}
\multicolumn{2}{c|}{Parameters} & 6.6M & - & - & - & - & 4.8M & 4.8M & 3.7M & 14.6M & 2.3M & 7.6M \\
\cline{1-13}
\multicolumn{2}{c|}{FLOPS} & 4.5G & - & - & - & - & -& - & - & 2.6G & - & 4.6G \\
\bottomrule
\end{tabular}
}
\vspace{-0.7cm}
\label{tab1}
\end{table*}

\vspace{-0.3cm}
\subsection{Universal Model Based Expert Specialization}

\begin{algorithm}[t]
\footnotesize
\caption{UMES}
\label{alg:umes}
\begin{algorithmic}[1]
\REQUIRE Training dataset $\mathcal{D}$, total epochs $E$, expert count $n$  
\ENSURE Specialized experts $\{f_i(\cdot)\}_{i=1}^n$ with params $\{\theta_i\}_{i=1}^n$
\STATE \textbf{Phase I: Universal} (epochs $1$ to $E/2$)
\STATE Initialize all experts with identical parameters $\theta_0$
\FOR{each batch $\mathbf{x} \in \mathcal{D}$}
    \STATE $o_1(\mathbf{x}) \gets \frac{1}{n}\sum_{i=1}^n f_i(\mathbf{x})$
    \STATE $\mathcal{L}_{\text{phaseI}} \gets \mathcal{L}_{\text{noise}}(\mathbf{x}) + \mathcal{L}_{\text{spk}}(\mathbf{x}|o_1(\cdot))$
    \STATE Update all experts with $\nabla_{\theta_0}\mathcal{L}_{\text{phaseI}}$
\ENDFOR
\STATE \textbf{Phase II: Specialization} (epochs $E/2+1$ to $E$)
\FOR{each batch $\mathbf{x} \in \mathcal{D}$}
    \STATE Compute gating weights $\{g_i(\mathbf{x})\}_{i=1}^n$
    \STATE $o_2(\mathbf{x}) \gets \sum_{i=1}^n g_i(\mathbf{x})f_i(\mathbf{x}|\theta_i)$
    \STATE $\mathcal{L}_{\text{phaseII}} \gets \mathcal{L}_{\text{phaseI}} + n\mathcal{L}_{\text{spk}}(\mathbf{x}|o_2(\cdot))$
    \STATE Update each expert with $\nabla_{\theta_i}\mathcal{L}_{\text{phaseII}}$
\ENDFOR
\RETURN $\{f_i(\cdot)\}_{i=1}^n$
\end{algorithmic}
\end{algorithm}

Inspired by the GMM-UBM \cite{b19}, we employ a Universal Model based Expert Specialization (UMES) strategy where expert networks first learn shared feature representations before specializing. As illustrated in Algorithm \ref{alg:umes}, UMES follows a two-phase curriculum to train a multi-expert system. Phase I builds a universal foundation: all experts start from identical parameters \(\theta_0\) and are updated uniformly through the averaged output \(o_1(\mathbf{x}) = \frac{1}{n}\sum_{i=1}^n f_i(\mathbf{x})\), equivalent to training a single shared model. Phase II enables specialization: experts inherit the universal parameters but are updated via differentiated gradients (visualized in Fig.\ref{fig1} (c)), where each expert's update is partially scaled by its noise-dependent gating weight \(g_i(\mathbf{x})\). This ensures experts develop noise-conditional expertise while maintaining robust performance through the preserved Phase I loss \(\mathcal{L}_{\text{phaseI}}\).\(\mathcal{L}_{\text{noise}}\) and \(\mathcal{L}_{\text{spk}}\) denote the cross-entropy losses of the noise and speaker classifiers, respectively. Multiplying \(n\) expert loss accelerates their specialization.
\vspace{-0.3cm}
\subsection{SNR-Decaying Curriculum Learning}
The SNR-Decaying Curriculum Learning (SDCL) enhances model learning efficiency by progressively reducing the Signal-to-Noise Ratio (SNR) in training data augmentation. It implements an easy-to-hard curriculum where the augmentation SNR is sampled from a truncated Gaussian distribution, with the distribution mean decaying exponentially across epochs:
\begin{equation}
\text{SNR} \sim \mathcal{N}_{\text{trunc}}(\mu_e, \sigma^2),\\
\mu_e = \exp\left(-k \cdot \frac{e}{E}\right)
\label{eq6}
\end{equation}
Here, $\mathcal{N}_{\text{trunc}}$ confines the SNR to effective ranges, $\mu_e$ controls the curriculum progression, $\sigma$ maintains diversity, $k$ is the decay rate, $e$ is the current epoch, and $E$ is the total number of epochs. Thus, SDCL avoids early exposure to extreme noise, enables gradual adaptation to diverse SNR levels, promotes expert specialization, and maintains training stability through controlled noise introduction. The truncated Gaussian provides stochastic variability within the curriculum framework.
\vspace{-0.3cm}
\section{Experiments}
\subsection{Data}
The experiments are conducted on the standard development and test sets of the Voxceleb1 dataset \cite{b20}. The development set contains 1211 speakers for training, while the test set consists of 37720 trials from 40 speakers for evaluation. To thoroughly assess system robustness, we construct noisy test conditions by augmenting the original clean utterances with MUSAN \cite{b21} and Nonspeech100 \cite{b22} at signal-to-noise ratios (SNRs) ranging from 0 to 20 dB in 5 dB increments. The MUSAN dataset includes ``babble'', ``music'' and ``noise''\footnotemark  as three distinct audio types,  which are partitioned into non-overlapping training and testing subsets to prevent data leakage, following the partition protocol in \cite{b11}. During model training, Each utterance undergoes online data augmentation through either additive noise mixing (with randomly selected samples from the MUSAN training set) or convolutional reverberation (using simulated room impulse responses). Training relied solely on noisy data with balanced noise types in augmentation. The remaining noise samples are reserved for synthesizing comprehensive evaluation sets that cover both in-domain and out-of-domain noise conditions.

\footnotetext{In the experimental tables, ``noise'' denotes the official MUSAN noise category, whereas in preceding sections it broadly includes all background interference (e.g., music, babble). }
\vspace{-0.3cm}
\subsection{Implementation Details}
The acoustic front-end extracts 80-dimensional log-mel filterbank features. Given its wide adoption, strong performance, and scalability, we use a ResNet with 32 initial channels as the baseline to validate the NCMoE framework. The speaker embedding dimension is 256. The model is trained for 150 epochs using SGD optimizer with the AAM-Softmax loss for speaker classification, employing mixed precision training to accelerate computation while maintaining stability. The proposed framework extends this baseline with four parallel residual branches in residual stage two, each following the identical structure as mentioned in Section II-2.1, while the noise classifier maintains a 32-channel initial width and adheres to the protocol of doubling channels during time-frequency downsampling. We treat reverberation as a structured noise and accordingly define four distinct augmentation categories as noise-type labels: ``noise'', ``babble'', ``music'' and ``reverberation''. The temperature factor $\gamma$  is empirically set at 0.1 , with the truncated Gaussian distribution bounded between 20 dB and 0 dB, and $\sigma$ empirically set at 0.2. The coefficient $k$ in Eq.\ref{eq6} is derived as 7.6 by logarithmic transformation to ensure that $\mu$ decays smoothly from 20 to 0.01 (approximately 0 for numerical stability) during training. The implementation is based on the Wespeaker \cite{b23} toolkit using 2 NVIDIA RTX 5070 Ti GPUs.

\vspace{-0.3cm}
\subsection{Results}

Table \ref{tab1} presents a comparative evaluation between the proposed method and existing approaches. Building upon a well-configured and competent baseline, our method delivers substantial performance gains through a novel framework of condition-specific subspace modeling. As further analyzed in Table \ref{tab2}, the method also demonstrates consistent gains under unseen noise conditions, confirming its enhanced cross-domain robustness. Notably, The framework incorporates a lightweight noise classifier to guide the selection among only lightweight, stage-2 experts. This design, powered by a single-expert activation mechanism, results in a minimal complexity overhead compared to the strong baseline. Consequently, it achieves a highly competitive balance between substantial performance gains and resource efficiency compared with the baseline. Furthermore, we compared the model complexity of relevant methods, and the reported FLOPs and parameters are estimated using publicly accessible data and open-source code. Our method is more complex but still within a comparable order of magnitude, while the actual growth in FLOPs is extremely limited compared with baselines. This is due to the inference approach with sparse activation of experts adopted in our method.

\begin{table}[t]
\footnotesize
\centering
\caption{Results (eer\%) on voxceleb1 test set with nonspeech100 data at various snrs.}
\vspace{-0.3cm}
\begin{tabular}{c|cccccc}
\toprule 
{SNR}& {Baseline} & {SEU-Net} & {Diff-sv} & NDAL & {NISRL} &{NCMoE} \\
\cline{1-7}
0 & 10.17 & \textbf{5.99} & 8.23 & 7.57 & 6.41 &6.27 \\
5 & 5.53 & 4.58 & 5.06 & 5.49 & 4.57 &\textbf{4.02}\\
10 & 3.79 & 3.74 & 3.85 & 4.03 & 3.55 &\textbf{3.05}\\
15 & 2.74 & 3.15 & 3.19 & 3.36 & 2.99 &\textbf{2.43}\\
20 & 2.36 & 2.87 & 2.89 & 2.99 & 2.75 &\textbf{2.19}\\ 
\cline{1-7}
Average & 4.92 & 4.07 & 4.65 & 4.97 & 4.05 & \textbf{3.59}\\
\bottomrule
\end{tabular}
\label{tab2}
\vspace{-0.4cm}
\end{table}

\vspace{-0.3cm}
\subsection{Further Analysis}
\begin{table}[t]
\footnotesize
\centering
\caption{Ablation study results (average eer\% across 5 snr levels) under varous synthetic noise conditions.}
\vspace{-0.3cm}
\begin{tabular}{c|cccc}
\toprule
Noise Type& NCMoE & w/o UMES & w/o NCER & w/o SDCL \\
\cline{1-5}
Babble & 3.96 & 9.50 & 4.02 & 4.05\\
Music & 2.87 & 7.14 & 3.00 & 2.98\\
Noise & 3.23 & 8.37 & 3.43 & 3.30\\
Nonspeech & 3.59 & 8.99 & 3.96 & 3.81\\
\cline{1-5}
Average & 3.41 & 6.80 & 3.60 & 3.54\\
\bottomrule
\end{tabular}
\label{tab3}
\vspace{-0.4cm}
\end{table}

\begin{table}[t!]
\footnotesize
\centering

\caption{Sensitivity analysis  (average eer\% across 5 snr levels) on hyper-parameters ($\gamma$ and $\sigma$).}
\vspace{-0.3cm}
\begin{tabular}{c|cc|cccc}
\toprule
Setting& $\gamma$=0.1 & $\gamma$=1.0 & $\sigma$=0.1 & $\sigma$=0.2 &$\sigma$=0.3 &$\sigma$=1.0 \\
\cline{1-7}
Average & 3.41 & 3.45 & 3.44 & 3.41 & 3.42 & 3.50\\
\bottomrule
\end{tabular}

\label{tab4}
\vspace{-0.4cm}
\end{table}

Our ablation study in Table \ref{tab3} demonstrates varying degrees of performance degradation when removing three key components: the UMES, NCER, and SDCL. The most significant performance drop occurs with the removal of the UMES strategy, indicating that excessive specialization among experts compromises discriminative speaker feature extraction due to their inherent shared patterns. Although eliminating the noise classification loss (w/o NCER) degrades performance, the model still maintains superiority over the baseline. Furthermore, the ablation of SDCL confirms that progressively reducing SNR during training effectively enhances model robustness. Moreover, Table \ref{tab4} summarizes the impact of key hyper-parameters, showing that the results do not fluctuate significantly with their variations.

\begin{figure}[t!]
\centerline{\includegraphics[width=0.85\linewidth]{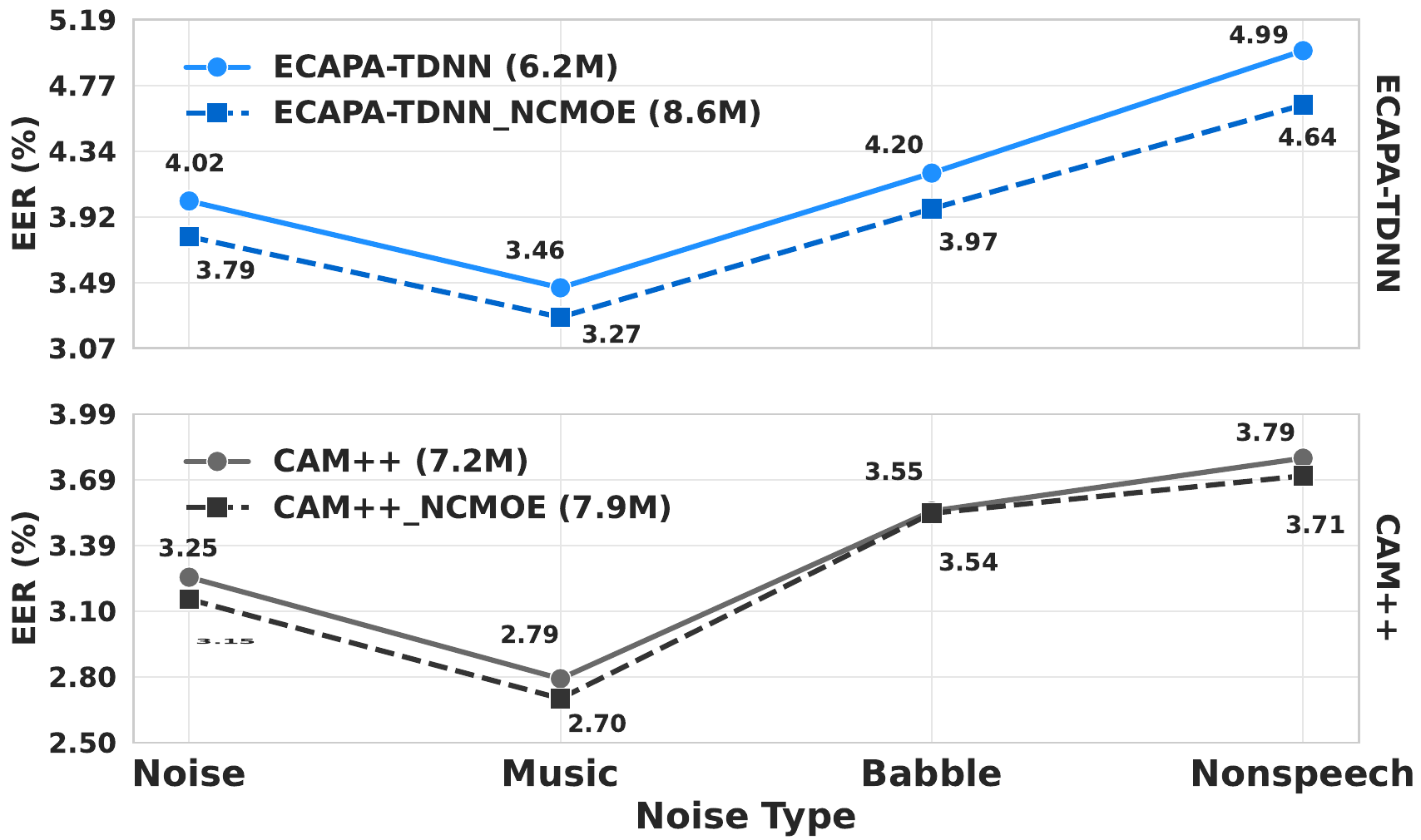}}
\vspace{-0.2cm}
\caption{EER performance comparison of NCMoE framework across different backbone networks.}
\label{fig2}
\end{figure}

To verify the generalization of our proposed method, we conducted experiments on two typical speaker recognition backbones (ECAPA-TDNN\cite{b5} and CAM++\cite{b27}). For a rational parameter-performance trade-off, only the second module of each backbone was extended into four expert networks. As shown in Fig. \ref{fig2}, though the model parameter count increased moderately, the average EER of each noise condition (across different SNRs) decreased consistently across all noise types. These results fully demonstrate the strong generality of our approach, which can effectively enhance the noise robustness of different backbone networks

\begin{figure}[t!]
\centerline{\includegraphics[width=1.0\linewidth]{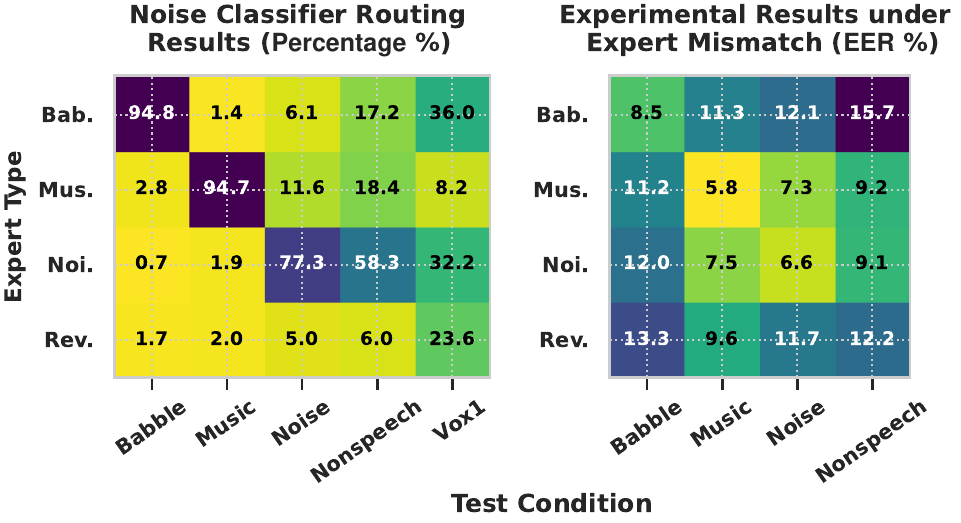}}
\vspace{-0.2cm}
\caption{Confusion Matrix of the routing result of the noise classifier and the EER of NCMoE under diverse training and testing audio conditions.}
\vspace{-0.5cm}
\label{fig3}
\end{figure}

To analysis the behavior of the noise classifier and the performance of individual experts in NCMoE, each expert (based on ECAPA-TDNN) was tested on data at 0 dB SNR. The resulting confusion matrix is shown in Fig.\ref{fig3}. Note that experts were independently initialized here for a clearer visualization. It shows that the noise classifier maintains high routing accuracy for babble and music, with moderate accuracy for noise. Out-of-domain data show distinct routing patterns, confirming the router's discriminative capability and validating the NCMoE framework. In addition, results show that performance degrades under most mismatched conditions. We attribute this to the distinct spectral characteristics of speech, music, and noise, each necessitating a specific noise suppression pattern. Mismatched expert routing applies the wrong suppression mode, leading to performance degradation.

\vspace{-0.3cm}
\section{Conclusion}
In this work, we propose a noise-conditioned mixture-of-experts framework for robust speaker verification. Unlike existing approaches that rely on a robust unified feature modeling space, our method decomposes the feature forwarding space into multiple noise-specific subspaces, enabling superior handling of diverse noise conditions. Extensive experiments validate the effectiveness of our proposed framework, demonstrating consistent performance improvements over the baseline. Future work will focus on more advanced mixture architectures and a refined noise classification front-end based on pre-trained models, which will enable the handling of broader and more detailed noise categorizations to further enhance the model's capabilities.

\bibliographystyle{IEEEtran}
\bibliography{refs.bib}
\end{document}